\documentclass[12pt,preprint]{aastex}

\slugcomment{Accepted by \apj}
\shorttitle{Magnetized Radiation-Dominated Disks}
\shortauthors{Turner, Stone, \& Sano}

\newcommand{\soses}{$\sigma/\sigma_\mathrm{es}$}
\newcommand{\dratio}{$\log\left<\left<
\rho_\mathrm{max}/\rho_\mathrm{min} \right>\right>$}
\newcommand{\lbobz}{\multicolumn{2}{c}{$\log\left|{\bf B}\right|/B_0$}}
\newcommand{\teoo}{$t_\mathrm{eqm}\Omega_0/2\pi$}
\newcommand{\divv}{{\bf\nabla\cdot v}}

\begin{document}
\title{Local Axisymmetric Simulations of Magneto-Rotational
Instability in Radiation-Dominated Accretion Disks}

\author{N. J. Turner\altaffilmark{1}, J. M. Stone\altaffilmark{1},
\& T. Sano\altaffilmark{1}}

\altaffiltext{1}{Astronomy Department, University of Maryland, College
Park, MD 20742; {\tt neal@astro.umd.edu}}

\begin{abstract}
We perform numerical simulations of magneto-rotational instability in
a local patch of accretion disk in which radiation pressure exceeds
gas pressure.  Such conditions may occur in the central regions of
disks surrounding compact objects in active galactic nuclei and
Galactic X-ray sources.  We assume axisymmetry, and neglect vertical
stratification.  The growth rates of the instability on initially
uniform magnetic fields are consistent with the linear analysis of
\citet{bs01}.  As is the case when radiation effects are neglected,
the non-linear development of the instability leads to transitory
turbulence when the initial magnetic field has no net vertical flux.
During the turbulent phase, angular momentum is transported outwards.
The Maxwell stress is a few times the Reynolds stress, and their sum
is about four times the mean pressure in the vertical component of the
magnetic field.  For magnetic pressure exceeding gas pressure,
turbulent fluctuations in the field produce density contrasts about
equal to the ratio of magnetic to gas pressure.  These are many times
larger than in the corresponding gas pressure dominated situation, and
may have profound implications for the steady-state vertical structure
of radiation-dominated disks.  Diffusion of radiation from compressed
regions damps turbulent motions, converting kinetic energy into photon
energy.
\end{abstract}


\section{INTRODUCTION}

The great decrease in specific angular momentum required to bring
material from typical galactic radii to the innermost stable orbit
round a central black hole indicates that accretion disks are likely
present in active galactic nuclei (AGN).  Direct evidence for such
disks includes the kinematics of maser emission observed on parsec and
sub-parsec scales in a few objects \citep{miyoshi95,gallimore96,gmh97}
and large asymmetries in the $6.4$~keV iron fluorescence line which
may be due to orbital motion in a general relativistic potential
\citep{tanaka95,fabian00}.  Some Galactic X-ray sources are associated
with collimated relativistic jets, suggesting disks occur around
stellar-mass compact objects also \citep{mr99}.  In a few cases, light
curves during eclipses by a companion star indicate an extended
structure about the compact object \citep{parmar86,ob97}.

One of the major uncertainties regarding disk accretion is how the
angular momentum is lost from incoming material.  A leading candidate
mechanism is the magneto-rotational instability (MRI), whose
importance for accretion disks was pointed out by \citet{bh91}.  The
MRI is driven by outward radial transport of angular momentum along
magnetic field lines sheared by differential orbital motion.  In most
previous studies of the instability, effects of radiation have been
neglected.  It has been found that provided dissipation of the field
is negligible and magnetic pressure is much less than gas pressure,
the fastest linear mode of the instability grows at near
three-quarters the orbital angular frequency $\Omega$, independent of
the field orientation.  This mode has a wavelength along the direction
parallel to the magnetic field of approximately $2\pi v_A/\Omega$,
where $v_A$ is the Alfv\'en speed \citep{bh98}.  In
magneto\-hydro\-dynamical (MHD) simulations of the non-linear
development with imposed axisymmetry, the instability develops from an
initially uniform vertical magnetic field into an
exponentially-growing two-channel flow.  From spatially-varying fields
with zero net vertical flux, short-lived turbulence is produced
\citep{hb92}.  In three-dimensional MHD calculations, the MRI leads to
outward angular momentum transport by long-lasting turbulence with
velocity fluctuations smaller than or comparable to the sound speed
\citep{hgb95,hgb96}.  An initially weak magnetic field grows to a
saturated level which is sustained for many orbits despite effects of
disk stratification \citep{bnst95,shgb96}.

A second major issue regarding disk accretion is how the released
gravitational energy is converted into radiation.  The energy released
in reaching the innermost stable orbit round a black hole is a
significant fraction of the rest energy of the incoming material.  In
parts of the accretion disk, radiation pressure may therefore greatly
exceed the ordinary gas pressure due to thermal motions of material
particles \citep{ss73}.  If the released energy is placed in the gas
through viscous or resistive dissipation, radiation may be produced by
thermal emission provided the absorption optical depth through the
disk is sufficient.  Alternatively if the released energy is placed in
turbulent motions, the kinetic energy might be converted directly into
photon energy by radiative damping.  This occurs when work is done on
the radiation in compressing a part of the flow, and photons diffuse
away from the region of higher energy density.  For linear compressive
MHD waves in the absence of rotation, damping may be rapid when the
wave period is longer than the time for photons to diffuse across the
wavelength \citep{ak98}.  During linear growth of the axisymmetric
MRI, radiative damping may be strong when in addition the pressure in
the azimuthal component of the magnetic field exceeds the gas pressure
\citep{bs01}.

In this paper we investigate the effects of radiation on the
non-linear development of the MRI.  As in the first such studies in
the gas pressure dominated regime \citep{hb91,hb92}, we assume
axisymmetry, neglect vertical stratification, and consider a small
patch in the interior of the disk.  The dependence of the magnetic
stresses on the gas and radiation pressures cannot be determined from
such axisymmetric calculations lacking sustained dynamo activity.
However, radiation diffusion and the inclusion of magnetic pressures
exceeding gas pressure may lead to new effects.  The equations solved
and numerical method used are described in \S~\ref{sec:method}, the
initial conditions and computational domain in \S~\ref{sec:ic}.
Results from the linear growth phase are compared against linear
analyses in \S~\ref{sec:linear}.  Effects of radiation on the channel
flow which develops from initially uniform vertical fields are
outlined in \S~\ref{sec:channel}.  Radiation effects in the decaying
turbulence on fields with zero net vertical flux are examined in
\S~\ref{sec:zeronet}, and the significance of the results for the
structure of accretion disks is discussed in \S~\ref{sec:discussion}.
A summary and conclusions are in \S~\ref{sec:conclusions}.

\section{EQUATIONS SOLVED AND NUMERICAL METHODS
\label{sec:method}}

The radiation magnetohydrodynamic equations solved are written in a
frame comoving with the radiating fluid, in cgs units.  Relativistic
effects are neglected.  Terms to order unity in $v/c$ are included,
and local thermodynamic equilibrium is assumed.  The equations are
\begin{equation}\label{eqn:cty}
{D\rho\over D t}+\rho\divv=0,
\end{equation}
\begin{equation}\label{eqn:gasmomentum}
\rho{D{\bf v}\over D t} = -{\bf\nabla}p
	+ {1\over 4\pi}({\bf\nabla\times B}){\bf\times B}
        + {1\over c}\chi\rho{\bf F},
\end{equation}
\begin{equation}\label{eqn:radenergy}
\rho{D\over D t}\left({E\over\rho}\right) =
	- {\bf\nabla\cdot F} - {\bf\nabla v}:\mathsf{P}
	+ \kappa\rho(4\pi B - c E),
\end{equation}
\begin{equation}\label{eqn:gasenergy}
\rho{D\over D t}\left({e\over\rho}\right) =
	- p\divv - \kappa\rho(4\pi B - c E),
\end{equation}
\begin{equation}\label{eqn:radmomentum}
{\bf F} = -{c\lambda\over\chi\rho}{\bf\nabla}E,
\end{equation}
and
\begin{equation}\label{eqn:induction}
{\partial{\bf B}\over\partial t} = {\bf\nabla\times}({\bf v\times B})
\end{equation}
\citep{mm84,smn92}.  Here the convective derivative $D/Dt$ is
equivalent to $\partial/\partial t + {\bf v\cdot\nabla}$.  The
dependent quantities $\rho$, $e$, ${\bf v}$, and $p$ are gas mass
density, energy density, velocity, and scalar isotropic pressure,
respectively, and ${\bf B}$ is magnetic field.  The radiation is
described by $E$, ${\bf F}$, and $\mathsf{P}$, the total
frequency-integrated radiation energy density, flux, and pressure
tensor, respectively.  Total opacity $\chi$ is the sum of absorption
$\kappa$ and scattering $\sigma$.  The opacities have units of
cross-section per unit mass, and are treated as independent of photon
frequency.  Emission proportional to the black-body rate $B=\sigma_B
T_g^4/\pi$ is assumed for the material component, where $\sigma_B$ is
the Boltzmann constant and $T_g=p\mu/({\cal R}\rho)$ the gas
temperature.  The gas constant is written ${\cal R}$, and the
dimensionless mean mass per particle $\mu$ is taken to be $0.6$.  An
ideal-gas equation of state is assumed, and the gas pressure is
related to the gas energy density by $p=(\gamma-1)e$ with
$\gamma=5/3$.  The temperature of the radiation field is computed by
$T_r=\left(E /a_R\right)^{1/4}$, where $a_R=4\sigma_B/c$.  Additional
terms in the gas momentum equation~\ref{eqn:gasmomentum} due to tidal
and rotating frame forces are discussed in \S~\ref{sec:ic}.

The equations are integrated using the ZEUS MHD code
\citep{sn92a,sn92b}.  Shocks are captured with a standard quadratic
artificial viscosity \citep{vr50}.  In ZEUS-2D, the azimuthal
component of the magnetic field is normally evolved as a passive
scalar, as described by \citet{sn92b}.  However, we have found that
this treatment results in a changing net azimuthal magnetic flux,
whereas the azimuthal component of the field evolution
equation~\ref{eqn:induction} integrated over the periodic shearing
sheet indicates that azimuthal flux ought to be conserved when radial
flux is zero \citep{hgb95}.  In trial calculations of the development
of magnetized turbulence in a shearing-sheet configuration, azimuthal
flux varied by up to 20\% when azimuthal field was evolved as a
scalar, and by a fraction of a percent when all components were
treated using the method of characteristics with constrained transport
(MOC-CT; Stone \& Norman 1992b).  The three components are advanced
together using MOC-CT in the remainder of the calculations reported
here.

The radiation terms in the equations are treated using the
flux-limited radiation diffusion (FLD) module described by
\citet{ts01}.  In the FLD approximation \citep{lp81}, neither a
radiation momentum equation nor a radiation transfer equation is
explicitly solved.  The flux is assumed to be ${\bf F}
=-D{\bf\nabla}E$, where the radiation diffusion coefficient
$D=c\lambda/(\chi\rho)$ incorporates a flux-limiter $0<\lambda\leq
1/3$.  The limiter is chosen according to the prescription of
\citet{lp81} so as to ensure the flux in optically-thin regions obeys
the causality constraint $|{\bf F}|\leq cE$.  The radiation pressure
tensor $\mathsf{P}=\mathsf{f}E$ is obtained using an Eddington tensor
$\mathsf{f}$ whose components depend on ${\bf\nabla}E$, rather than on
full solutions of the radiation transfer equation.  In the majority of
the calculations discussed in this paper, each grid zone is
optically-thick.  Under these conditions the flux-limiter is very
close to $1/3$, the radiation pressure is isotropic and described by
the scalar $P$, and the FLD method is equivalent to a two-temperature
diffusion approximation.

\section{GRID AND INITIAL CONDITIONS
\label{sec:ic}}

We use the shearing-sheet approximation \citep{hb92}.  A small patch
of the disk is represented in local corotating Cartesian coordinates
$(x, y, z)$.  At time $t$ the origin of the local coordinates lies at
$(R_0, \Omega_0 t, 0)$ in global inertial cylindrical coordinates $(R,
\phi, z)$.  The local coordinates rotate with the Keplerian orbital
frequency $\Omega_0 = (G M/R_0^3)^{1/2}$ appropriate for a central
mass $M$ and gravitational constant $G$.  Symmetry along the $y$-axis
or direction of orbital motion is assumed.  The radial component of
the gravitational force due to the central mass is included in the
$x$-component of the gas momentum equation~\ref{eqn:gasmomentum}
through a tidal term $3\rho\Omega_0^2 x$ on the right-hand side.  The
Coriolis force due to the rotation of the frame is incorporated
through a term $-2\rho\Omega_0{\bf\hat{z}\times v}$, where
${\bf\hat{z}}$ is a unit vector along the rotation axis.  The vertical
component of the gravitational force is not included.  Initially the
orbital or $y$-component of the velocity varies linearly with $x$ in
an expansion of the Keplerian rotation curve about $R_0$, while the
$x$- and $z$-components are small or zero.  Mass density and gas and
radiation energy densities are uniform, and gas and radiation are in
thermal equilibrium.  The $x$- and $z$- boundaries are taken to be
periodic, except that $y$-velocity is offset across the $x$-boundaries
according to the differential orbital motion.  The grid zones are
square and evenly-spaced.

\begin{figure}
\epsscale{0.65}
\plotone{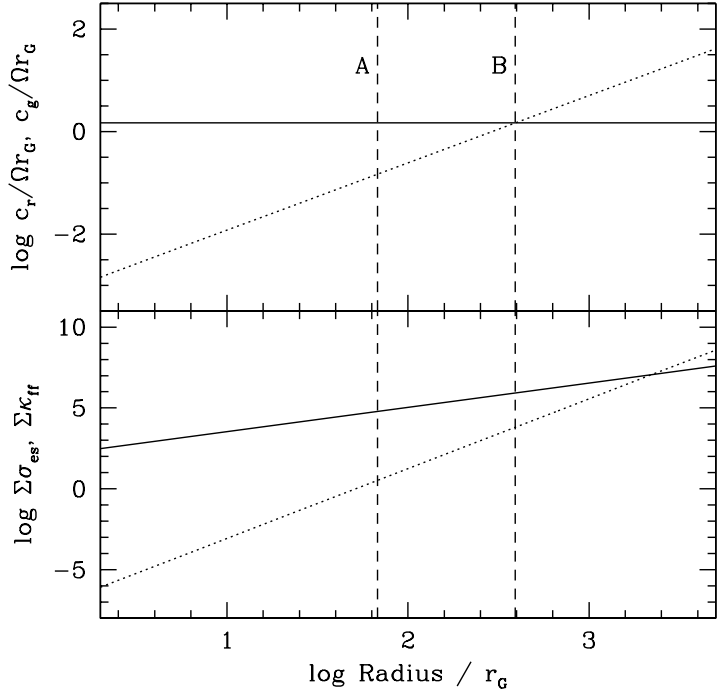}
\caption{Standard initial conditions are taken from this
Shakura-Sunyaev model disk accreting into a $10^8 M_\odot$ black hole
at 10\% of the Eddington rate, with luminous efficiency 0.1 and
$\alpha=0.01$.  The horizontal axis is distance from the center of
mass in gravitational radii $r_G=GM/c^2$.  In the upper panel are
shown the disk half-thicknesses in units of $r_G$, corresponding to
support by radiation pressure (solid) and gas pressure (dotted).  In
the lower panel are the total optical depths due to electron
scattering (solid) and free-free absorption (dotted).  Dashed vertical
lines indicate the radii where the local midplane radiation acoustic
speed is ten times greater than the gas acoustic speed (labeled~A) and
equal to the gas acoustic speed (labeled~B).  Initial conditions for
most of the simulations are taken from one of these two locations.
\label{fig:ic}}
\end{figure}

Steady-state conditions might occur in an accretion disk when gains
and losses of both magnetic and radiation energy balance.  In the
present axisymmetric calculations, sustained dynamo action is not
expected and surface losses are ignored, so the outcome is largely
fixed by the starting conditions.  We explore several initial field
strengths and geometries, and set the gas and radiation pressures and
the diffusion rate by choosing initial mass and energy densities from
a standard time-steady, optically-thick viscous accretion disk model
\citep{ss73}.  In this model, energy loss by radiative diffusion
through the disk surfaces is assumed to balance vertically-integrated
dissipation at each radius, and accretion is due to an effective
viscous shear stress proportional to local radiation plus gas pressure
$P+p$ with constant of proportionality $\alpha$.  The half-thickness
of the disk $H$ is determined by vertical hydrostatic balance, and is
independent of radius $R$ in the radiation-dominated region because
the vertical components of the forces due to gravity and radiation
both scale as $R^{-3}$.  Estimates of the surface density $\Sigma$ and
midplane radiation acoustic speed $c_r=(\frac{4}{3}P/\rho)^{1/2}$, gas
acoustic speed $c_g=(\gamma p/\rho)^{1/2}$, and opacities are obtained
neglecting relativistic corrections and assuming matter and radiation
are in local thermodynamic equilibrium.  The opacities used are due to
electron scattering, $\sigma_\mathrm{es}=0.4$~cm$^2$~g$^{-1}$, and
free-free absorption, $\kappa_\mathrm{ff}=10^{52}\rho^{9/2}
e^{-7/2}$~cm$^2$~g$^{-1}$.  For the range of parameters considered
here, Comptonization and bound-free opacities have little effect on
interior structure of an $\alpha$-disk \citep{hbka01}.  Results are
shown in figure~\ref{fig:ic} for a central mass $M = 10^8 M_\odot$,
accretion rate 10\% of the Eddington rate $\dot M_E=2.65\times 10^{-9}
(M/M_\odot) \eta^{-1} M_\odot$~yr$^{-1}$ with luminous efficiency
$\eta=0.1$, and $\alpha$ parameter 0.01.  Starting values for the
majority of the calculations are taken from the radius marked~A, where
radiation acoustic speed is ten times gas acoustic speed.  This
location lies at 67.8 gravitational radii $r_G=GM/c^2$.  The Keplerian
orbital speed here is $0.12\,c$, the radiation acoustic speed is 2.2\%
of the Keplerian speed, and radiation pressure is 125 times gas
pressure.  The power law index in the equation of state of gas plus
radiation is therefore close to $4/3$.  For some additional
calculations, initial conditions are taken from location~B at
392$r_G$, where the two acoustic speeds are equal.  The radiation
pressure is $1.25$ times the gas pressure, so that this location is
near the outer edge of the radiation-dominated region.  Under these
conditions the power law index in the combined equation of state is
expected to be $1.4$ \citep[p.\ 320]{mm84}, intermediate between those
for radiation and gas separately.  In numerical calculations with
location~B initial conditions, acoustic vibrations in the absence of
radiation diffusion have speeds consistent with the expected equation
of state.  The initial conditions taken from the midplane at
locations~A and~B are listed in table~\ref{tab:ic}.  The effective
absorption optical depth of the disk depends on the distances traveled
by photons scattered multiple times between emission and escape, and
is
\begin{equation}\label{eqn:taueff}
\tau^* = \Sigma \sqrt{\sigma\kappa}.
\end{equation}
As may be seen from table~\ref{tab:ic}, the initial disk model is
effectively optically thick to both free-free absorption and electron
scattering at the two locations considered.  Local absorption and
emission of photons bring gas and radiation into equilibrium in a time
\begin{equation}\label{eqn:teqm}
t_\mathrm{eqm}\approx{e\over c\kappa\rho E}.
\end{equation}
We have verified by numerical integration of the coupled gas and
radiation energy equations~\ref{eqn:radenergy} and~\ref{eqn:gasenergy}
with ${\bf v}={\bf F}={\bf 0}$ that equation~\ref{eqn:teqm}
approximately describes the damping of gas and radiation pressure
perturbations at locations~A and~B.  As indicated in
table~\ref{tab:ic}, the equilibration time is orders of magnitude
shorter than the orbital period at both locations.  When the physical
equilibration time is shorter than the numerical timestep, the
implicit differencing scheme used in the FLD module remains stable,
and gas and radiation are brought into thermal equilibrium in a few
timesteps \citep{ts01}.  Finally, we have checked that the initial
conditions are dynamically stable when no magnetic field is present.
Applied radial and azimuthal velocity perturbations lead to epicyclic
oscillations at frequency $\Omega_0$.

\begin{deluxetable}{rll}
\tablewidth{0pt}
\tablecaption{Initial conditions\label{tab:ic}}
\tablehead{Location & A & B }
\startdata
$R/r_G$                 &$67.8$                 &$392$                  \\
$c_r/c_g$               &$10$                   &$1$                    \\
$P/p$                   &$125$                  &$1.25$                 \\
$\tau_\mathrm{es}$      &$6.2\times 10^4$       &$8.6\times 10^5$       \\
$\tau^*_\mathrm{ff}$    &$4.5\times 10^2$       &$7.5\times 10^4$       \\
\teoo                   &$1.2\times 10^{-6}$    &$4.6\times 10^{-9}$    \\
$\rho$/g~cm$^{-3}$      &$2.89\times 10^{-9}$   &$4.02\times 10^{-8}$   \\
$T_g=T_r$/K		&$2.71\times 10^5$	&$1.40\times 10^5$	\\
$e$/erg cm$^{-3}$       &$1.63\times 10^5$      &$1.17\times 10^6$      \\
$E$/erg cm$^{-3}$       &$4.07\times 10^7$      &$2.93\times 10^6$      \\
\enddata
\end{deluxetable}

The values of $\rho$, $e$, and $E$ from the disk model listed in
table~\ref{tab:ic} are inserted as initial conditions in radiation MHD
simulations described in the remainder of this article.  These
simulations include no viscous accretion stress, and no viscous
heating apart from the standard artificial viscosity described in
\S~\ref{sec:method}.  The only source terms in the azimuthal component
of the equation of motion~\ref{eqn:gasmomentum} are the Lorentz and
Coriolis forces.

\section{COMPARISON WITH LINEAR ANALYSES
\label{sec:linear}}

In this section we examine the linear growth of the MRI.  Results from
numerical calculations are compared against analytic estimates for
magnetic fields which are initially vertical, vertical plus azimuthal,
and radial.  These comparisons may reveal shortcomings of our
numerical techniques.

\citet{bs01} considered the case where radiation pressure is
important, absorption opacity is zero, and the initial magnetic field
has a vertical and possibly an azimuthal component.  When
stratification is neglected, the dispersion relation obtained by an
axisymmetric WKB analysis is their equation~39.  The unstable modes of
the MRI are found to be unchanged by radiation diffusion when the
field is initially vertical.  Numerical results for this geometry are
discussed in \S~\ref{sec:uniformbz}.  When the initial azimuthal
magnetic pressure exceeds the gas pressure, the growth rates are
reduced if the time for radiation to diffuse across the vertical
wavelength is shorter than the orbital period.  Numerical results in
several such cases are presented in \S~\ref{sec:uniformbzbp}.

For initially radial magnetic fields, no linear analysis including
effects of radiation is presently available.  In the case where gas
pressure is greater than magnetic pressure and radiation effects are
negligible, the general dispersion relation from \citet{bh91}
indicates that the fastest-growing incompressible axisymmetric mode
has radial wavelength approximately $2\pi v_A/\Omega_0$, vertical
wavelength asymptotically zero, and growth rate near
$\frac{3}{4}\Omega_0$.  Some numerical results with and without
radiation are discussed in \S~\ref{sec:uniformbr}.

Growth rates in the simulations below are measured by computing the
time rate of change in the Fourier power spectrum components of the
$x$-velocity distribution, about one orbital period before the end of
the linear growth stage.  The velocity distribution is sampled every
0.1~orbital period.

\subsection{Uniform Vertical Magnetic Field
\label{sec:uniformbz}}

Initial conditions and orbital frequency for these calculations are
extracted from location~A in the disk model discussed in
\S~\ref{sec:ic}.  The magnetic field is initially vertical, with
strength chosen so the Alfv\'en speed $v_A=\left|{\bf B}
\right|/\sqrt{4\pi\rho}$ is equal to the gas acoustic speed.  The
height and width of the computational domain are set to the vertical
wavelength of the mode to be examined, and the domain is divided into
$64^2$ zones.  The $x$-velocity is perturbed with a sinusoidal
variation in $z$ having the wavelength of the mode.  Its initial
amplitude is smaller than the Keplerian speed at $R_0$ by a factor
$10^8$, and smaller than the radiation acoustic speed by a factor
$2.2\times 10^6$.  Both electron scattering and free-free absorption
opacities are included.

Modes with wavelengths $2/3$, $1$, and $2$ times the characteristic
MRI wavelength $2\pi v_A/\Omega_0$ have growth rates differing by less
than 1\% from those predicted using the \citet{bs01} dispersion
relation, their equation~39.  The three modes grow at similar rates
with the free-free absorption opacity set to zero, and with the
scattering opacity instead reduced a factor 100; under these latter
conditions, radiation diffuses across the domain in less than an
orbit.  These results support the conclusion of \citet{bs01} that on
initially vertical fields in a scattering medium, the spectrum of
unstable modes of the MRI is unaffected by radiation diffusion.  The
small absorption opacity also has little effect on the modes.

\subsection{Uniform Vertical Plus Azimuthal Magnetic Field
\label{sec:uniformbzbp}}

For the first set of calculations in this section, the initial
conditions are identical to those in \S~\ref{sec:uniformbz} except
that a large azimuthal component is added to the magnetic field.  The
total Alfv\'en speed is equal to the radiation acoustic speed, and is
ten times the vertical Alfv\'en speed $v_{Az}=B_z/\sqrt{4\pi\rho}$.
The \citet{bs01} dispersion relation indicates that the fastest
axisymmetric MRI mode in this situation has wavelength near $2\pi
v_{Az}/\Omega_0$.  Modes with wavelengths $0.78$, $1.16$, and $2.33$
times this characteristic value are examined.

\begin{figure}
\epsscale{0.65}
\plotone{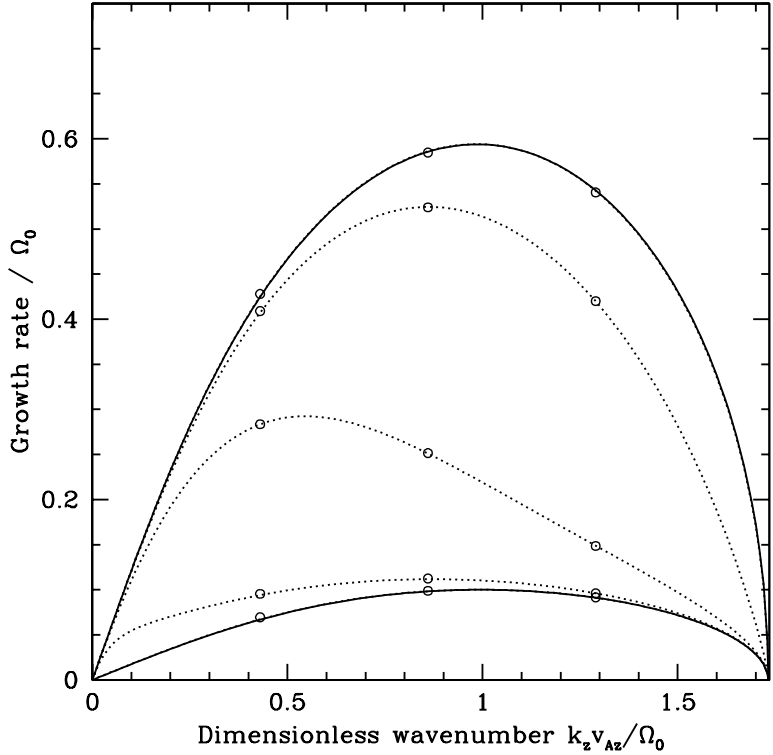}
\caption{Linear growth rate of the MRI versus wavenumber, on an
initially uniform magnetic field.  The field has azimuthal and
vertical components, the vertical being 10\% of the magnitude.  Growth
rate is plotted against the product of vertical wavenumber $k_z$ and
vertical Alfv\'en speed $v_{Az}$.  Both axes are labeled in units of
the orbital frequency $\Omega_0$.  Axisymmetry is assumed, and the
radial wavenumber $k_x$ is zero.  Solid curves: limits in which
diffusion is very slow (upper) and very fast (lower).  Dotted curves:
solutions of the \citet{bs01} dispersion relation including radiation,
with diffusion rate increasing from top to bottom.  Total Alfv\'en
speed is equal to radiation acoustic speed, and is ten times gas
acoustic speed.  The dotted curve peaking near growth rate
$0.5\Omega_0$ corresponds to the diffusion rate at location~A in the
model disk discussed in \S~\ref{sec:ic}.  The uppermost dotted curve,
which overlaps the upper solid curve, is for a diffusion rate 100
times slower.  The lower curves are for diffusion rates 10, 100, and
$10^4$ times faster.  The lowest dotted curve overlaps the lower solid
curve.  Circles indicate growth rates measured in the linear phases of
corresponding numerical simulations.
\label{fig:uniformbzbp}}
\end{figure}

In the other calculations in this section, the diffusion rate is
varied by choosing initial conditions from time-steady viscous
accretion disk models with different values of the $\alpha$ parameter.
For each diffusion rate, the expected dependence of growth rate on
wavenumber is shown by a dotted curve in figure~\ref{fig:uniformbzbp}.
Growth rates measured in the simulations are indicated by open
circles.  Overall, the agreement between the linear analysis and the
numerical results is within about 1\%.  Use of the same numerical
method in studying the non-linear development therefore seems
worthwhile.

\subsection{Uniform Radial Magnetic Field
\label{sec:uniformbr}}

We first check growth rates in a simulation without radiation effects
against those predicted using the dispersion relation from
\citet{bh91}.  Solution of their equation~2.17 indicates that on
radial field the fastest-growing mode has zero vertical wavelength.
To locate the fastest mode present on the grid, random perturbations
are applied to the $x$- and $z$-velocities.  Each component has
probability distribution uniform between $+$ and $-10^{-8}$ times the
Keplerian speed at $R_0$.  Initial density is as in
\S~\ref{sec:uniformbz}, and gas pressure is set to the sum of the gas
and radiation pressures in \S~\ref{sec:uniformbz}.  Alfv\'en speed is
set to 10\% of its value there, so that $\beta=p/(\left|{\bf B}
\right|^2/8\pi)$ is $15\,120$.  The grid consists of $64\times
64$~zones, and the domain height and width are twice the
characteristic wavelength $2\pi v_A/\Omega_0$.

The mode which grows fastest in the linear stage has two wavelengths
across the width of the box and 32 in the height, and grows at
$0.726\Omega_0$.  The growth rate expected for this mode from the
linear analysis is $0.7478\Omega_0$.  Although the vertical wavelength
is two grid zones and the mode is grossly under-resolved, the measured
growth rate is only slightly less than the expected value.

Diffusion of radiation is likely to be rapid in configurations with
such short vertical wavelengths.  If radiation pressure support is
lost, the velocity pattern of the perturbation may acquire a non-zero
divergence when shear acting on radial magnetic field produces an
azimuthal field exceeding equipartition with the gas.  Such
compressible motions are associated with reduced growth rates when
radiation effects are unimportant \citep{bb94,ko00}.

In a second calculation, with the same initial magnetic field as above
but including a radiation energy density as in \S~\ref{sec:uniformbz}
so the ratio of gas to magnetic pressure $\beta=120$, we find that the
measured growth rate declines below $0.1\Omega_0$ over several orbits
as azimuthal magnetic pressure grows larger than gas pressure.  To
allow measurement of the weak-field linear growth rate, an initial
magnetic field ten times smaller is selected for a third calculation.
The size of the domain is reduced to twice the new characteristic MRI
wavelength.  In this case $\beta$ is initially $12\,000$, and
equipartition between field and gas is not reached for several orbits.
The fastest-growing mode observed has two wavelengths in the domain
width and 28 in the height, and grows at $0.721\Omega_0$.  The
differences between the second and third calculations indicate that
radiation diffusion can greatly reduce the growth rate of the MRI on
initially radial magnetic field, when the magnetic pressure exceeds
the gas pressure.

\section{NON-LINEAR EVOLUTION ON UNIFORM VERTICAL MAGNETIC FIELD
\label{sec:channel}}

In the non-linear regime, the axisymmetric MRI leads to development of
exponentially-growing channel solutions when the initial field has a
net vertical flux and radiative and resistive effects are slight
\citep{hb92,gx94}.  Any smaller-scale features undergo an inverse
cascade until the domain is filled by an inward-moving and an
outward-moving channel, separated by regions of strong radial and
azimuthal magnetic field.  We have carried out calculations of this
phenomenon with and without radiation effects, using identical
initially uniform vertical fields.  Initial mass and energy densities
for the case with radiation are as at location~A in the disk model of
\S~\ref{sec:ic}, and the field strength is such that Alfv\'en speed
equals gas acoustic speed.  For the case without radiation, the
initial gas energy density is chosen to give the same total pressure.
The width and height of the domain are twice the characteristic
wavelength of the MRI.  A radial velocity perturbation independent of
$x$ is applied.  Its vertical wavelength is equal to the
characteristic wavelength, and its amplitude is $10^{-4}$ times the
Keplerian orbital speed at $R_0$.

\begin{figure}
\epsscale{0.65}
\plotone{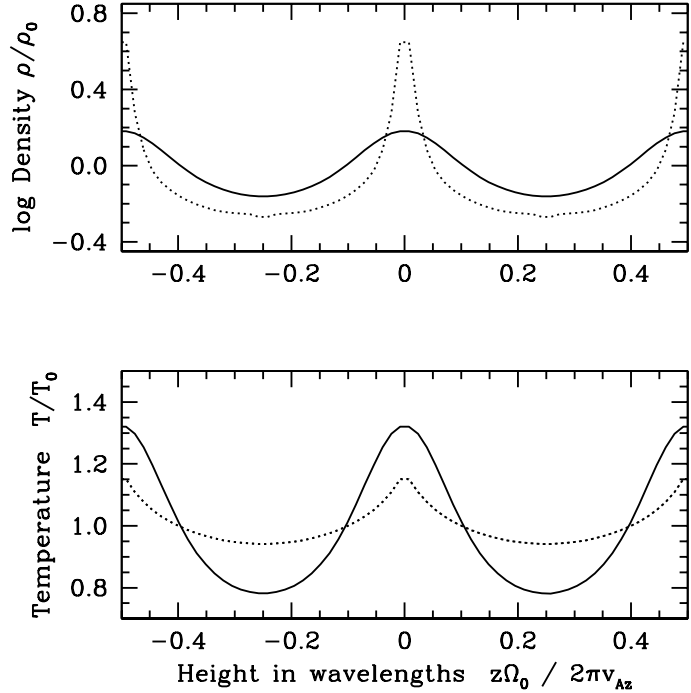}
\caption{Non-linear development of the MRI from an initially uniform
vertical magnetic field, as discussed in \S~\ref{sec:channel}.
Variation of density (upper) and temperature (lower) with height at
$x=0$ is shown at the time when the pressure of the magnetic field
perturbation first exceeds initial gas plus radiation pressure.  Only
the central wavelength of the disturbance is plotted.  Results from a
calculation including radiation pressure are shown by dotted curves.
Those from a calculation without radiation, having initial gas
pressure increased to give the same total pressure, are shown by solid
curves.  There are 64~grid zones per wavelength in both cases.
\label{fig:channel}}
\end{figure}

Results at the start of the non-linear stage are shown in
figure~\ref{fig:channel}.  In the case with radiation, diffusion from
the compressed layers into the lower-density, magnetically-dominated
layers leads to faster compression and a weaker temperature contrast.
Oppositely-directed radial and azimuthal magnetic fields are brought
into a single grid zone in the compressed layer somewhat earlier in
the case with radiation, and rapid numerical reconnection of the field
follows.

\section{NON-LINEAR EVOLUTION ON FIELD WITH ZERO NET VERTICAL FLUX
\label{sec:zeronet}}

In previous axisymmetric calculations without radiation, the linear
development of the MRI on a magnetic field with zero net vertical flux
is followed by a period of turbulence.  The fluctuations die away over
several orbits as vertical field of opposing signs is brought together
and lost from the calculation through numerical dissipation
\citep{hb92}.  In this section we describe the results of a series of
simulations in which the strengths of the radiation effects are varied
to determine how they affect the transitory turbulence.  Throughout,
the domain average of a quantity $Q$ is indicated $\left<Q\right>$.
Time-averaged values are indicated by a second pair of angle braces
$\left<\left<Q\right>\right>$.

\subsection{Canonical Simulation
\label{sec:canonical}}

The uniform initial values of the density and gas and radiation energy
densities for the canonical simulation are taken from location~A as
described in \S~\ref{sec:ic}.  The initial magnetic field is given a
large azimuthal component because this configuration leads to strong
effects of radiation diffusion during the linear growth.  The strength
of the initial field $B_0$ is made independent of position and the
radial component is set to zero, so that there are no unbalanced
forces.  The field direction varies sinusoidally with~$x$.  At grid
center $x=0$, the field is inclined $\theta=+5.7^\circ$ from
azimuthal, so that the vertical component $B_z=0.1 B_0$.  At the
boundaries $x=\pm L/2$, the field is inclined $-5.7^\circ$, and at
$x=\pm L/4$, it is entirely azimuthal.  The domain has height and
width $L$.  The components of the field are
\begin{equation}\label{eqn:initialbz}
B_z=B_0\sin\theta \cos\left(2\pi x/L\right),
\end{equation}
\begin{equation}\label{eqn:initialby}
B_y=\left(B_0^2-B_z^2\right)^{1/2},
\end{equation}
and
\begin{equation}\label{eqn:initialbx}
B_x=0,
\end{equation}
with $\sin\theta=0.1$.  The azimuthal component $B_y$ is everywhere
positive.  Field strength $B_0$ is chosen so Alfv\'en speed is equal
to radiation acoustic speed and is ten times gas acoustic speed.  The
pressures due to radiation, magnetic field, and gas are in the ratio
$125:83.3:1$.  Each zone is given small random $x$- and
$z$-velocities.  The probability distribution for each component of
the velocity is linear between $-10^{-4}$ and $10^{-4}$ times the
Keplerian speed at the central radius $R_0$.  The domain size $L$ is
set to twice the characteristic wavelength of the axisymmetric MRI
$2\pi v_{Az}/\Omega_0$ measured at grid center, and the grid consists
of $128^2$ zones.  The domain height is 51\% of the thickness of the
disk model from which the initial conditions were selected.  In the
initial state, the time for radiation acoustic waves to cross the
domain is 0.2~orbits, and photons diffuse across the domain in $L^2/D
= 50$~orbits.

\begin{figure}
\epsscale{0.65}
\plotone{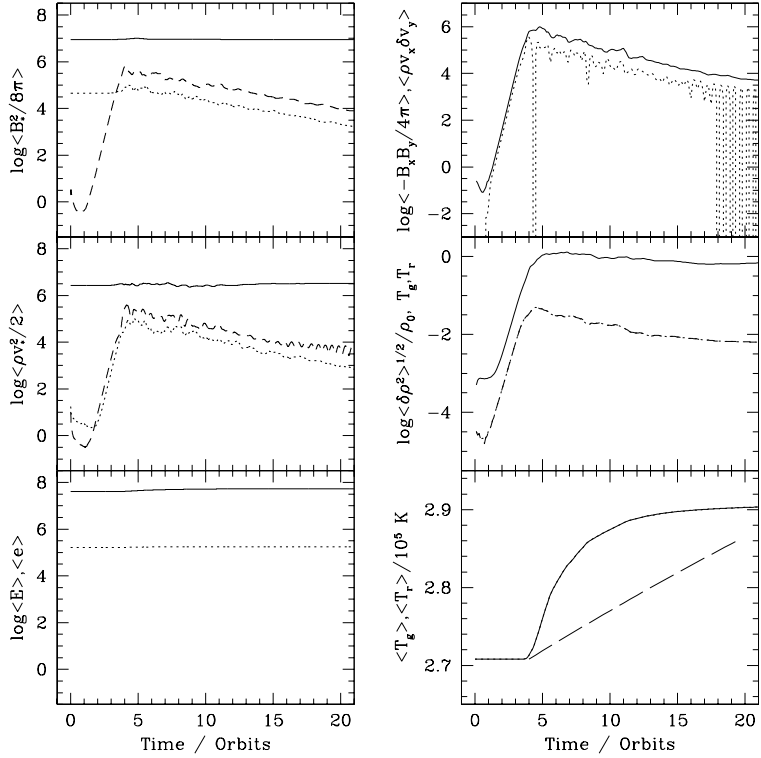}
\caption{Time history of domain-averaged quantities in the canonical
simulation described in \S~\ref{sec:canonical}.  Magnetic and kinetic
energy densities measured in the local frame are shown in the top two
panels on the left.  The energies in the $x$-, $y$-, and
$z$-components are indicated by dashed, solid, and dotted lines,
respectively.  Radiation (solid) and gas (dotted) energy densities are
plotted on the same scale in the lower left panel.  At upper right are
the Maxwell (solid) and Reynolds (dotted) accretion stresses.  In the
center right panel are the RMS fluctuations in density (solid), gas
temperature (dotted) and radiation temperature (dashed).  At lower
right are the temperatures of the radiation (solid) and gas (dotted,
hidden by solid line).  The heating which would result from the energy
release rate assumed in the initial disk model at location~A is
indicated by a long-dashed line.
\label{fig:time}}
\end{figure}

After about one orbit during which the energy in the initial velocity
perturbations is redistributed among gas, radiation, magnetic, and
kinetic energies, the simulation proceeds through three stages, as
shown in figure~\ref{fig:time}.  During the linear stage, covering the
remainder of the first 4.1~orbits, a mode with vertical wavelength
equal to the characteristic MRI wavelength grows exponentially at
about half the orbital frequency.  This mode has nulls at $x=\pm L/4$,
where the vertical component of the field is zero.  The linear stage
ends when the radial field resulting from the instability is
comparable in strength to the initial vertical field.  Inward and
outward moving flows then collide, and the simulation enters the
turbulent stage.  We choose to place the beginning of the turbulent
stage at the time when the energy in the radial component of the field
first peaks.  The turbulence decreases in intensity as the energy in
the vertical component of the field falls due to numerical losses.
After about 17~orbits, the Maxwell stress drops below 1\% of its peak
value, and the Reynolds stress oscillates about zero with twice the
epicyclic frequency.  The flow remains in this third, almost quiescent
stage until the end of the simulation at 30~orbits.  In the other
simulations described below, the flow passes through the same three
stages.  The durations of the linear and turbulent stages are
different in each case.  In each simulation we focus on the period
from 2 to 6 orbits after the start of the turbulent stage, when
transients reflecting the linear development have been largely erased,
and the fluctuations remain strong.

\subsection{Turbulent Fluctuations
\label{sec:fluctuations}}

The density variations during the turbulent stage in the canonical
simulation are larger than in similar gas pressure dominated
calculations.  The RMS fluctuation averaged from 2 to 6 orbits after
the onset of turbulence is 110\% of the mean density
(figure~\ref{fig:time}, center right panel).  On average, half the
mass is contained in the 18\% of the volume where densities are
greatest.  Throughout the same period the largest density in the
domain is about 200 times the smallest.  Overdense regions form and
disperse often.  Typically a clump is present about one orbit before
being destroyed either in a collision with another clump, or in
passing through a region of shear in the turbulence.

Density fluctuations of similar magnitude are seen in versions of the
canonical simulation with the domain size extended to 2.5 and 4
characteristic MRI wavelengths, and $160^2$ and $256^2$ zones,
respectively.  These calculations are listed in
table~\ref{tab:fluctuations} as R2.5w and R4w.  From these results we
infer the large densities are not due solely to the limited number of
linearly unstable modes present in the canonical domain.  There are
three linear modes with $k_x=0$ whose vertical wavelengths evenly
divide the canonical domain height, whereas six such modes are present
in the double-size calculation R4w.

Similar density fluctuations are also seen in a double-resolution
version of the canonical simulation R1d2 having $256^2$ zones.  The
highest and lowest densities are about the same in the canonical and
double-resolution calculations, and the densest regions have similar
physical sizes such that radiation diffuses out of them in
approximately an orbital period.  It is likely that the densest
regions are adequately resolved in the canonical calculation.

\begin{deluxetable}{cclrr@{.}lll}
\tabletypesize{\small}
\tablewidth{0pt}
\tablecaption{Simulations at location~A with strong azimuthal field.
\label{tab:fluctuations}}
\tablehead{
\colhead{Label}    & \colhead{Name}    &
\colhead{$L/(2H)$} & \colhead{Zones}   &
\lbobz             & \colhead{\soses}  &
\colhead{\dratio}  }
\startdata
R1   &Canonical              &$0.51$  &$128^2$& $0$&$0$ &$1$    &$2.30$   \\
R2   &                       &$0.16$  &$128^2$&$-0$&$5$ &$1$    &$1.32$   \\
R3   &                       &$0.051$ &$128^2$&$-1$&$0$ &$1$    &$0.46$   \\
R4   &                       &$0.016$ &$128^2$&$-1$&$5$ &$1$    &$0.081$  \\
R5   &                       &$0.0051$&$128^2$&$-2$&$0$ &$1$    &$0.0061$ \\
R2.5w&2.5-wavelength         &$0.64$  &$160^2$& $0$&$0$ &$1$    &$2.25$   \\
R4w  &4-wavelength           &$1.02$  &$256^2$& $0$&$0$ &$1$    &$1.79$   \\
R1d2 &Double-resolution      &$0.51$  &$256^2$& $0$&$0$ &$1$    &$2.20$   \\
R1t  &Timesteps/10           &$0.51$  &$128^2$& $0$&$0$ &$1$    &$2.22$   \\
R1b  &Negative-$B_y$         &$0.51$  &$128^2$& $0$&$0$ &$1$    &$1.95$   \\
R1v  &Artificial viscosity/10&$0.51$  &$128^2$& $0$&$0$ &$1$    &$1.99$   \\
R1s  &Very high opacity      &$0.51$  &$128^2$& $0$&$0$ &$10^4$ &$0.287$  \\
R6   &High-opacity           &$0.51$  &$128^2$& $0$&$0$ &$100$  &$0.276$  \\
R7   &                       &$0.16$  &$128^2$&$-0$&$5$ &$100$  &$0.150$  \\
R8   &                       &$0.051$ &$128^2$&$-1$&$0$ &$100$  &$0.113$  \\
R9   &                       &$0.016$ &$128^2$&$-1$&$5$ &$100$  &$0.039$  \\
R10  &                       &$0.0051$&$128^2$&$-2$&$0$ &$100$  &$0.0061$ \\
N11  &Same gas pressure      &$0.51$  &$128^2$& $0$&$0$ &\ldots &$1.90$   \\
N12  &                       &$0.16$  &$128^2$&$-0$&$5$ &\ldots &$0.90$   \\
N13  &                       &$0.051$ &$128^2$&$-1$&$0$ &\ldots &$0.32$   \\
N14  &                       &$0.016$ &$128^2$&$-1$&$5$ &\ldots &$0.200$  \\
N15  &                       &$0.0051$&$128^2$&$-2$&$0$ &\ldots &$0.085$  \\
N16  &Same total pressure    &$0.51$  &$128^2$& $0$&$0$ &\ldots &$0.280$  \\
N17  &                       &$0.16$  &$128^2$&$-0$&$5$ &\ldots &$0.193$  \\
N18  &                       &$0.051$ &$128^2$&$-1$&$0$ &\ldots &$0.082$  \\
N19  &                       &$0.016$ &$128^2$&$-1$&$5$ &\ldots &$0.023$  \\
N20  &                       &$0.0051$&$128^2$&$-2$&$0$ &\ldots &$0.0078$ \\
\enddata
\tablecomments{Calculations including radiation effects have labels
with prefix R.  Calculations without radiation have prefix N.  The
initial magnetic geometry is described in \S~\ref{sec:canonical}.}
\end{deluxetable}

All of these results may be compared against two versions of the
canonical calculation without radiation.  In the first, labeled N11 in
table~\ref{tab:fluctuations}, the initial gas pressure is unchanged
from the canonical calculation and is 1.2\% of the magnetic pressure.
Linear growth rates in this situation are as on the lower solid curve
in figure~\ref{fig:uniformbzbp}.  During the turbulent stage in run
N11, mean gas pressure increases monotonically owing to compression
and artificial viscous heating accompanying shocks.  The gas pressure
time-averaged from 2 to 6 orbits after the start of turbulence is 14\%
of the corresponding magnetic pressure.  The time-averaged RMS density
fluctuation in this case is 68\%, and the mean ratio of the largest
density to the smallest is 80.  These are almost as large as in the
canonical simulation.

In the second calculation without radiation, the radiation pressure is
replaced by an equal amount of additional gas pressure, so initially
the gas pressure is 1.5 times the magnetic pressure.  This calculation
is labeled N16 in table~\ref{tab:fluctuations}.  Linear growth rates
under these conditions lie near the upper solid curve in
figure~\ref{fig:uniformbzbp}.  During the turbulent stage in run N16,
the ratio of time-averaged gas to magnetic pressure is $1.69$.  This
is comparable to the ratio $1.79$ in the canonical simulation, of
time-averaged gas plus radiation pressure to magnetic pressure.  The
time-averaged RMS density fluctuation in N16 is 8.4\%, and the typical
density range is a factor two, about two orders of magnitude less than
in the canonical simulation.

\begin{figure}
\vspace*{5mm}
\epsscale{0.65}
\plotone{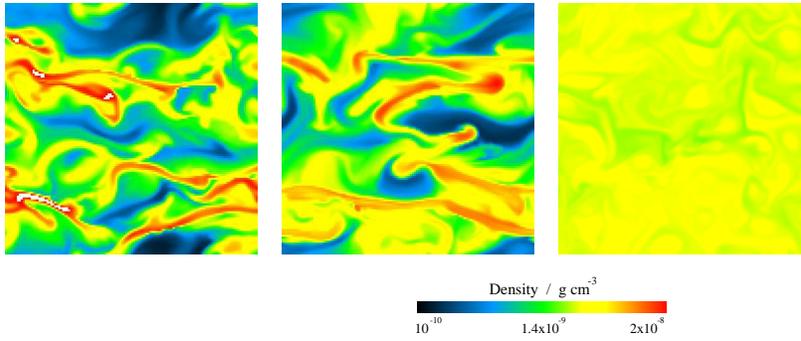}
\caption{Snapshots of the density distribution three orbits after
turbulence begins, in the canonical simulation R1 (left) and
calculations without radiation having the same initial gas pressure
(N11, center) and the same initial total pressure (N16, right).  A
common logarithmic scale is used.  The radial coordinate $x$ increases
to the right in each panel, and $z$ increases upwards.
\label{fig:density}}
\end{figure}

\begin{figure}
\epsscale{0.65}
\plotone{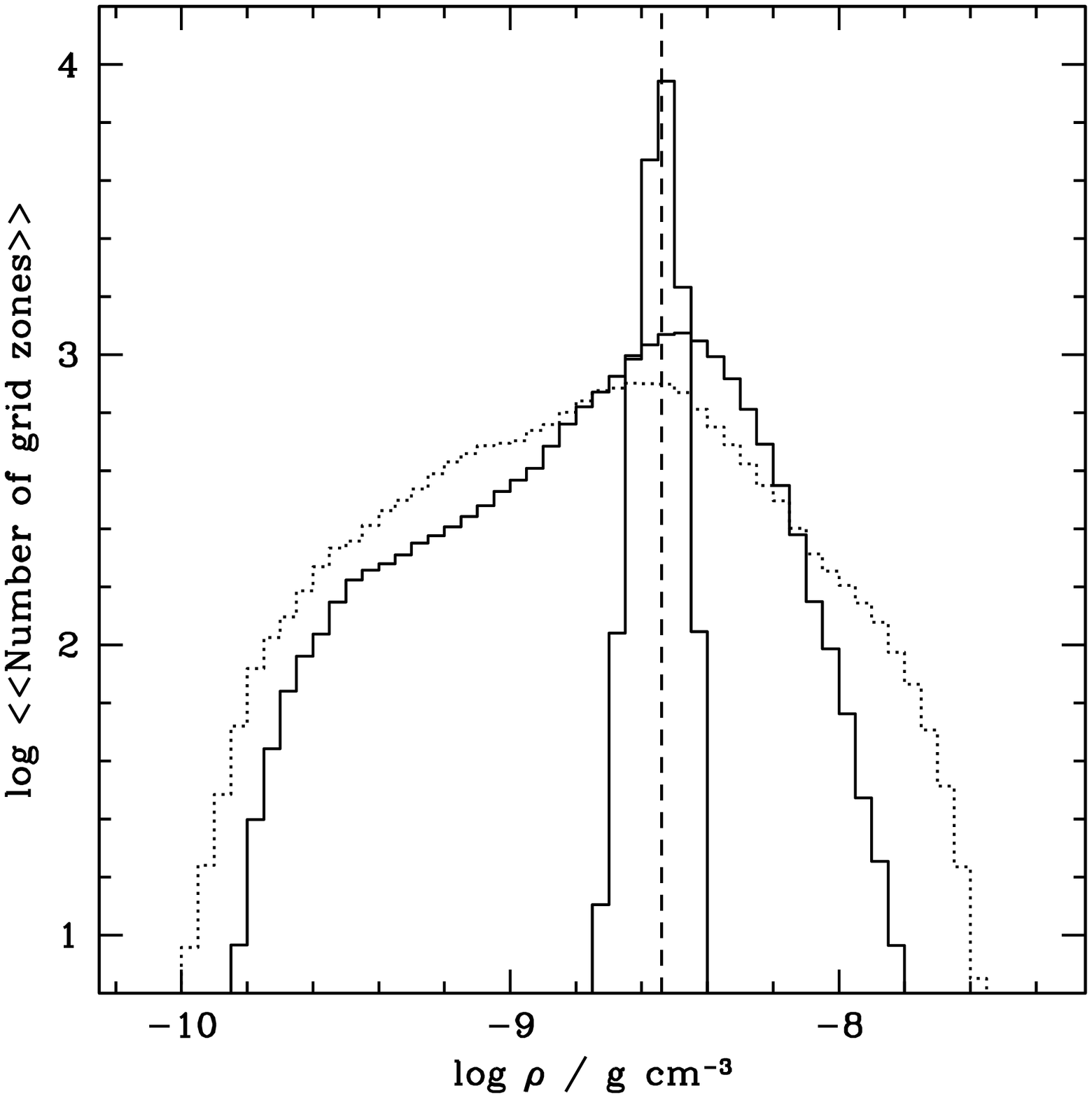}
\caption{Density histograms averaged over the period from 2 to 6
orbits after the onset of turbulence.  The canonical calculation R1 is
indicated by a dotted line.  Calculations without radiation, on the
same grid and with the same initial magnetic field, are shown by solid
lines.  For the broader solid histogram, from calculation N11, the
initial gas pressure is as in the canonical case.  For the taller,
narrower solid histogram, from calculation N16, initial gas pressure
is larger so that total pressure matches the canonical calculation.
The mean density for all three calculations is indicated by a vertical
dashed line.
\label{fig:histogram}}
\end{figure}

Snapshots of the density distributions in the three calculations
3~orbits after the start of the turbulent stage are shown in
figure~\ref{fig:density}.  A similar relationship between the density
fluctuations in the three calculations extends over the whole
averaging period.  Histograms of the density, time-averaged from 2~to
6~orbits after the onset of turbulence, are shown in
figure~\ref{fig:histogram}.  The two calculations with large density
fluctuations show quite different linear growth rates.  The canonical
simulation enters the turbulent stage after 4.1~orbits, while the
version with the same initial gas pressure but without radiation
reaches a similar amplitude only after 18.8~orbits.  The linear growth
rate does not appear to be a good predictor of the strength of the
density fluctuations in the turbulent stage in these axisymmetric
calculations.

Total pressure is almost uniform in the canonical simulation.  Regions
of higher density and gas and radiation pressures correspond to
regions of lower magnetic pressure.  The time-averaged RMS fluctuation
in magnetic pressure is 13~times larger than the corresponding
fluctuation in gas pressure, and is roughly equal to the fluctuation
in gas plus radiation pressure.  In the two calculations N11 and N16
without radiation, the gas and magnetic pressure fluctuations are
roughly equal.  Apparently in the canonical calculation, diffusion
does not completely decouple gas and radiation over a dynamical time,
and radiation pressure gradients provide some support against magnetic
forces.

The fractional RMS gas temperature fluctuations in the canonical
simulation are 30 to 100 times smaller than the density fluctuations
(figure~\ref{fig:time}, center right panel).  By contrast, in both of
the calculations without radiation, temperature fluctuations are
comparable to density fluctuations, as expected for adiabatic
evolution.  In the canonical simulation the radiation energy density
is more smoothly distributed than the gas energy density, owing to
diffusion of the radiation with respect to the gas.  Diffusion is
faster in the turbulent stage than initially, due to the concentration
of most of the opacity in small dense regions.  The turbulence and the
distribution of radiation evolve on the orbital timescale, while gas
and radiation reach thermal equilibrium in a few timesteps.  As there
are about $10^4$ timesteps per orbit, gas and radiation remain close
to equilibrium throughout.  The smaller gas temperature fluctuations
in the canonical calculation are due to radiation diffusion combined
with rapid thermal equilibration.  Additional mechanisms for exchange
of energy between material particles and radiation, such as Compton
scattering, are likely to have little further effect on the
temperatures.  The actual equilibration time due to free-free opacity
(table~\ref{tab:ic}) is initially about 100 times shorter than the
timestep.  Its smallest local value during the turbulent stage is
14~times shorter than its initial value.  Its largest is 33~times
longer than the initial value.  Temperature fluctuations are similar
in a version of the canonical simulation~R1t with Courant number
reduced a factor~10, so the timesteps are closer to the true
matter-radiation equilibration time.

The dense regions in the canonical calculation have lower magnetic
pressure than their surroundings primarily due to weaker azimuthal
fields.  The radial fields in dense regions on average point towards
larger~$x$, such that differential rotation produces azimuthal fields
opposing the background azimuthal field.  The time average of the
density-weighted mean radial field is $\left<\left<\rho B_x \right> /
\left<\rho\right>\right> = 1337$~G.  This is comparable to the initial
amplitude of the vertical component of the field, $B_0 \sin\theta =
1507$~G.  In a version of the canonical calculation R1b having the
sign of the initial azimuthal field reversed, fields in dense regions
preferentially point towards smaller~$x$, and the density-weighted
mean radial field is $-876$~G.  Radial magnetic field of opposite
signs is partially segregated in R1 and R1b, with fields of one sign
found mostly in dense regions, and those of the other sign found more
often in low-density regions.  The mean radial field in overdense
regions, the spatial anti-correlation between gas and magnetic
pressure fluctuations, the near-matches between magnetic pressure
fluctuations and gas plus radiation pressure fluctuations, and the
Maxwell stresses larger than the Reynolds stresses
(figure~\ref{fig:time}, upper right panel) are all consistent with a
picture in which the fluctuations in the gas and radiation are caused
by the magnetic fluctuations.

To investigate the importance of field strength for the density
fluctuations, versions of the canonical simulation are carried out
with initial magnetic fields weaker by factors of $10^{0.5}$, 10,
$10^{1.5}$, and 100.  These are labeled R2 through R5 in
table~\ref{tab:fluctuations}.  The magnetic pressure is less than the
gas pressure for the last three of these cases.  In the weakest-field
calculation~R5, the ratio of gas to magnetic pressure $\beta=120$.
For each run, the height and width of the domain are reduced by the
same factor as the magnetic field, so initially there are 64~zones per
characteristic MRI wavelength as in the canonical simulation.

\begin{figure}
\epsscale{0.65}
\plotone{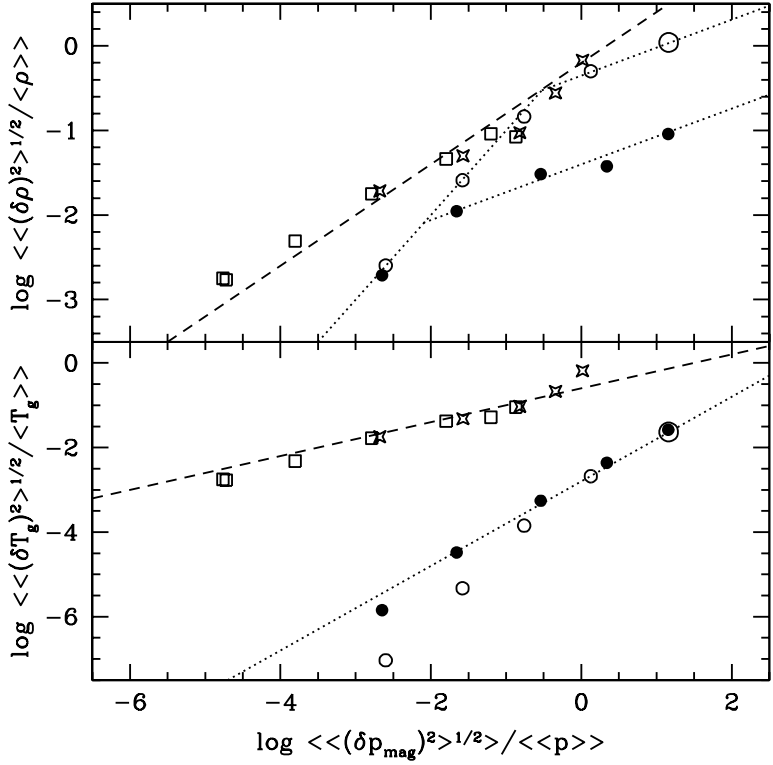}
\caption{Dependence of the RMS fluctuations in density (upper) and gas
temperature (lower) on the RMS magnetic pressure fluctuation in
calculations with various initial field strengths.  The magnetic
pressure fluctuation is measured relative to the gas pressure.  The
fluctuations and gas pressure are time-averaged over the period from 2
to 6 orbits after the onset of turbulence.  Calculations R1-R5 with
radiation and the standard opacities are marked by open circles.  The
largest open circle represents the canonical simulation~R1.
Calculations R6-R10 with scattering opacity 100~times larger are
indicated by filled circles.  Runs without radiation N11-N15 having
the same initial gas pressure are marked by stars.  Those N16-N20
having the same initial total pressure are marked by squares.  In the
upper panel, the steepest dotted line indicates the unit slope
expected for isothermal fluctuations, and the other two dotted lines
have slope~0.33.  Dashed lines indicate the slopes 0.6 (upper) and 0.4
(lower) expected for adiabatic fluctuations.  The dotted line in the
lower panel has unit slope.
\label{fig:fluctuations}}
\end{figure}

Time-averaged RMS fluctuations in the resulting turbulence are marked
in figure~\ref{fig:fluctuations} by open circles.  Among the three
weakest-field calculations, fluctuations in density and pressure are
proportional, as expected for isothermal perturbations.  Also, the
magnetic pressure fluctuations are approximately equal to the gas
pressure fluctuations in each case, and the radiation pressure
fluctuations are much smaller than both.  For the canonical and $|{\bf
B}|=B_0/10^{0.5}$ calculations, the density fluctuation is
proportional approximately to the pressure fluctuation raised to the
power~$0.33$.

Effects of diffusion on the density fluctuations are explored using a
parallel series of calculations R6 through R10, with scattering
opacity 100~times larger.  Results from these runs are shown in
figure~\ref{fig:fluctuations} by filled circles.  The break between
slopes unity and~$0.33$ is at a characteristic MRI wavelength 10~times
smaller than in the standard-opacity series.  In both cases, the break
occurs where the initial characteristic wavelength is approximately
equal to the distance radiation diffuses in an orbit.  The effective
equation of state is isothermal for fluctuations weaker than the break
point, and markedly stiffer for stronger fluctuations.  The large
density fluctuations observed in the canonical calculation require
both a large ratio of magnetic to gas pressure, and a sufficiently
high rate of radiation diffusion.

\subsection{Heating Mechanisms
\label{sec:heating}}

Radiation energy density increases with time in the canonical
calculation, as shown in figure~\ref{fig:time}.  The mean heating rate
from 2 to 6 orbits after the onset of turbulence is similar to the
dissipation rate $\frac{3}{4} \Omega_0 \alpha (P+p)$ assumed in the
steady-state viscous disk model of \S~\ref{sec:ic}.  In this section
we consider the sources of the heating in the canonical calculation.

The overall energy balance for gas plus radiation is obtained by
adding equations~\ref{eqn:radenergy} and~\ref{eqn:gasenergy} and
integrating over the domain.  Since the boundaries are periodic and
the transport terms and ${\bf\nabla\cdot F}$ term conserve total
energy, these terms integrate to zero, leaving
\begin{equation}\label{eqn:heating}
{\partial\over\partial t}\left<E+e\right> = -\left<p\divv\right>
 - \left<{\bf\nabla v}:\mathsf{P}\right>.
\end{equation}
Under optically-thick conditions as in the canonical run, the
radiation pressure tensor is isotropic and the final term reduces to
$-\left<P\divv\right>$.  When radiation pressure is much larger than
gas pressure, the radiation compression term dominates.  Throughout
the turbulent phase in the canonical calculation, the compression
heating rate given by equation~\ref{eqn:heating} is about an order of
magnitude larger than the domain-averaged rate of heating due to the
shock-capturing artificial viscosity.  Integrated from~2 to~6 orbits
after turbulence begins, the compression terms and artificial
viscosity increase the total heat content $\left<E+e\right>$ by 10.4\%
and 0.8\% of the initial value, respectively.  The viscous heating
rate assumed in the $\alpha$-disk model of \S~\ref{sec:ic} corresponds
to an increase in heat content by 6.3\% over the same period.
Compression heating in the canonical run is fast enough to be
considered as a possible mechanism for converting released
gravitational energy to photons.

\begin{figure}
\epsscale{0.65}
\plotone{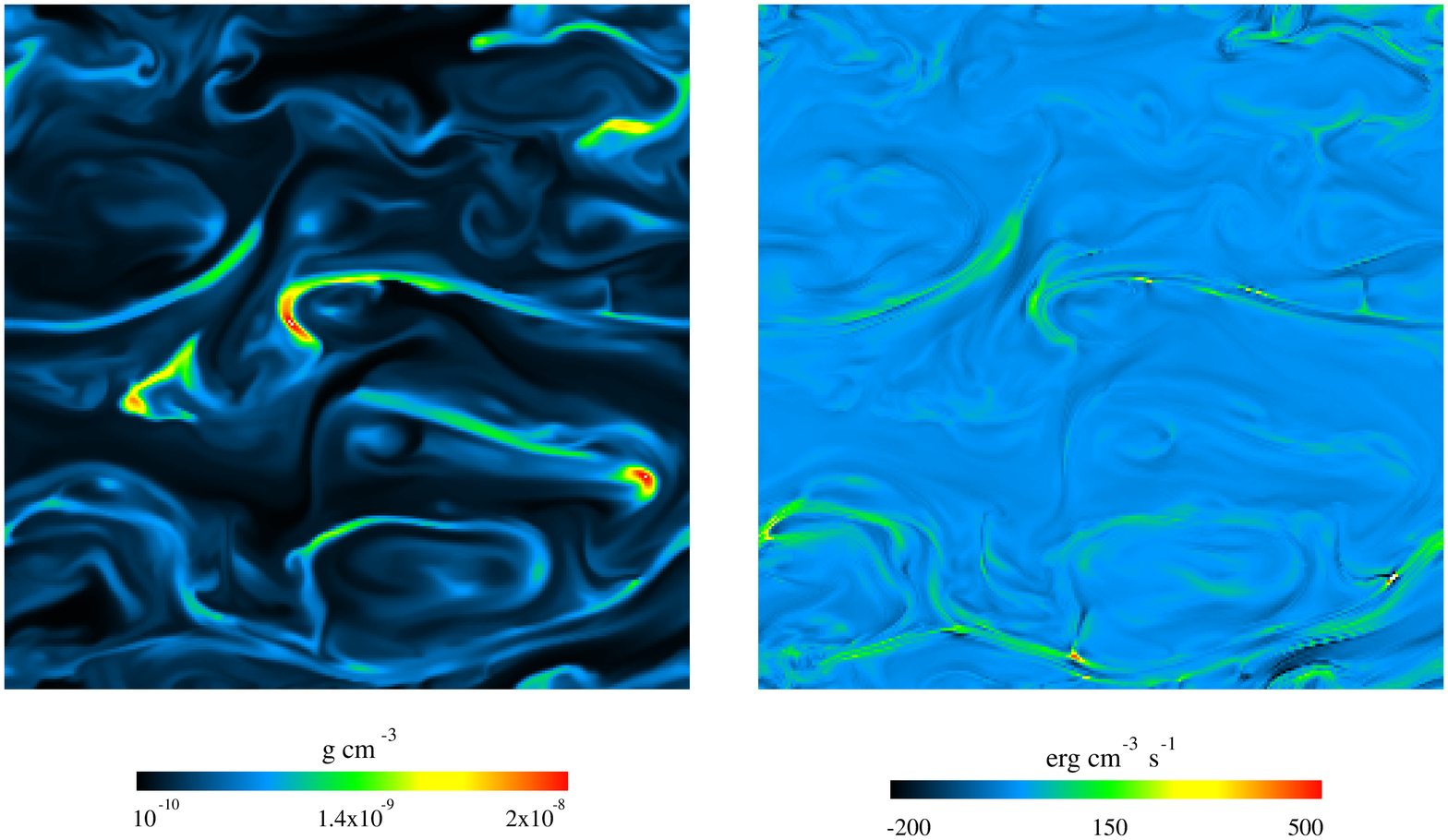}
\caption{Snapshots of the density (left) and compression heating rate
$-(P+p)\divv$ (right), three orbits after turbulence begins in the
double-resolution version of the canonical simulation, R1d2.  The
scale for the density is logarithmic, and that for the heating rate is
linear.  Compression heating is most commonly rapid in layers
surrounding regions of high density.
\label{fig:pdv}}
\end{figure}

In the double-resolution version of the canonical calculation~R1d2,
the corresponding increases due to the compression terms and the
artificial viscosity are 10.5\% and 1.6\%.  In a version~R1v with
artificial viscosity ten times smaller, they are 6.8\% and 0.098\%.
The amount of compression heating changes little with resolution and
with the artificial viscosity.  In these cases as in the canonical
run, the rate of heating due to $-(P+p)\divv$ is typically greatest
where material is compressed onto the boundaries of dense regions, as
shown in figure~\ref{fig:pdv}.  In contrast, in the calculation~N11
with the same initial gas pressure without radiation, the
time-integrated contributions of compression and artificial viscosity
are 4.1\% and 1.4\% of the initial total energy in the canonical
calculation, respectively.  In the calculation~N16 with the same total
pressure without radiation, the contributions are 1.6\% and 0.7\%.  In
the two calculations without radiation, the compression heating rate
is most often large in the interiors of dense regions.

Pressure is more strongly anti-correlated with velocity divergence in
the canonical calculation than in the versions lacking radiation
effects.  Anti-correlation indicates gas plus radiation pressure is
larger during compression than during expansion.  Redistribution of
the internal energy occurs by diffusion of radiation after
compression.  The difference in average total pressure $P+p$ between
regions of converging and diverging flow is $19.2\times
10^5$~dyn~cm$^{-2}$ in calculation~R1, $3.36\times 10^5$~dyn~cm$^{-2}$
in calculation~N11, and $8.82\times 10^5$~dyn~cm$^{-2}$ in
calculation~N16.

\begin{figure}
\vspace*{5mm}
\epsscale{0.65}
\plotone{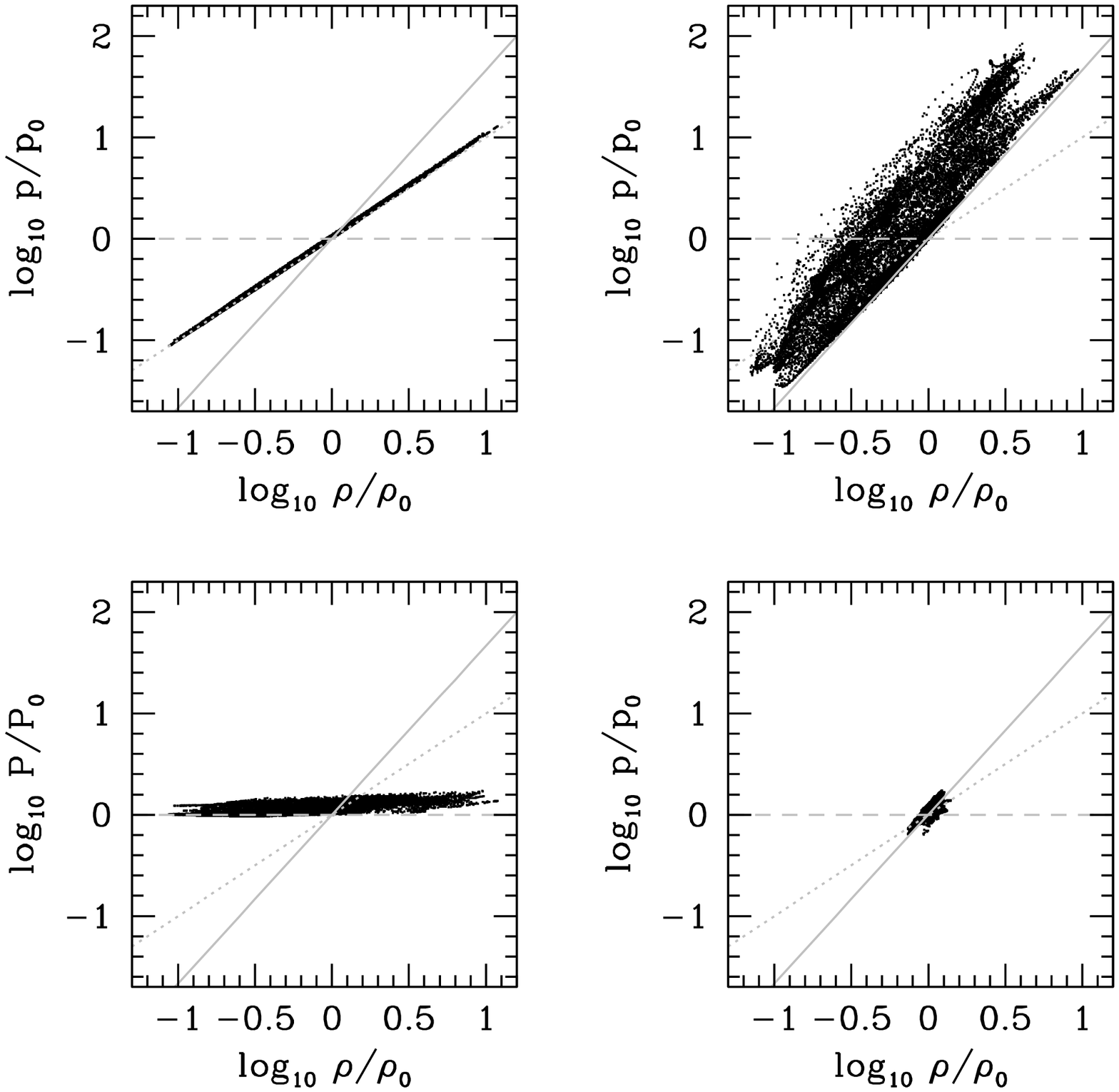}
\caption{Gas pressure (upper left) and radiation pressure (lower left)
versus density in the canonical calculation~R1; and gas pressure
versus density in the calculations without radiation having the same
initial gas pressure (N11, upper right) and the same initial total
pressure (N16, lower right).  Conditions in all $128^2$ zones are
shown 2~orbits after the onset of turbulence.  Each quantity is
plotted on a logarithmic scale, in units of its initial value.  Grey
lines indicate the initial pressures (dashed) and the paths for
disturbances of the initial conditions which are adiabatic (solid) and
isothermal (dotted).
\label{fig:eos}}
\end{figure}

The importance of diffusion in separating radiation from gas is shown
by the almost-uniform radiation pressure in the canonical calculation,
and by the pressure-density relations plotted in figure~\ref{fig:eos}.
In the canonical run, radiation pressure is nearly independent of
density except at the highest densities, while gas pressure is
proportional to density.  In the calculations~N11 and~N16 without
radiation, the relationships between pressure and density are
consistent with adiabatic evolution together with artificial viscous
heating.

The time-integrated compression heating depends strongly on magnetic
pressure.  It exceeds the heating due to artificial viscosity only in
the calculations R1 through R3 with total magnetic pressure similar to
or larger than gas pressure.  Compression heating is also less when
the fluid is more nearly opaque.  In a version of the canonical
simulation~R1s with a larger scattering opacity $\sigma=10^4
\sigma_\mathrm{es}$, the integrated compression heating is 1.6\% and
artificial viscous heating is 0.7\% of the initial total energy
density.  These are about the same as in the calculation~N16 without
radiation having the same initial total pressure.  We conclude that
radiation damping is associated with density excursions, and occurs
under the conditions outlined at the end of \S~\ref{sec:fluctuations}.
For rapid compression heating, magnetic pressure must be at least
comparable to gas pressure, and the radiation diffusion time not much
longer than the time for evolution of the turbulence.

\subsection{Field Geometry and Accretion Stress
\label{sec:geometry}}

Here we examine the effects of different initial magnetic field
orientations on the accretion stress.  In the canonical calculation,
the total stress time-averaged from 2 to 6 orbits after the start of
turbulence is 1.6 times the value $\alpha(P+p)$ assumed in the initial
disk model of \S~\ref{sec:ic}.  The stress which accompanies the early
part of the turbulent stage is roughly consistent with the starting
conditions used.

In a version of the canonical run R21, the azimuthal component of the
field is reduced to the same amplitude as the vertical component.  The
total magnitude of the field is reduced to $|{\bf B}|=B_0/10$, and the
field angle $\theta$ in
equations~\ref{eqn:initialbz}-\ref{eqn:initialbx} is set to
$90^\circ$.  As in the canonical run, the energy in the azimuthal
component of the field remains roughly constant during the turbulent
stage, while the energy in the vertical and radial components declines
through numerical dissipation.  Averaged from 2 to 6 orbits after the
onset of turbulence, the pressures in the components of the magnetic
field are in the ratio
$\left<B_x^2\right>:\left<B_y^2\right>:\left<B_z^2\right> = 4.8:248:1$
in the canonical simulation, and $2.1:7.1:1$ in the simulation with
weaker azimuthal field.  In these axisymmetric calculations, the
initial magnetic geometry affects the orientation of the field in the
turbulent stage.

For both of these calculations~R1 and~R21, the domain- and
time-averaged magnetic accretion stress $\left<\left< -B_x B_y/4\pi
\right>\right>$ is a few times the average pressure in the vertical
component of the magnetic field.  The stress due to correlated
hydrodynamic fluctuations $\left<\left< \rho v_x\delta v_y
\right>\right>$ is a few times smaller than the magnetic stress.
Similar ratios between the two stresses and the $z$-magnetic pressure
occur in weaker azimuthal field, $\theta=90^\circ$ versions of
calculations~R3 and~R5.  These are listed as~R23 and~R25 in
table~\ref{tab:geometry}.  Field strength and domain size are smaller
than in~R21 by factors of 10 and 100, respectively.  Similar ratios of
the stresses are also found in each of the simulations with and
without radiation listed in table~\ref{tab:fluctuations}.  The
accretion stresses in these calculations vary little with the
azimuthal field, the gas pressure, and the radiation pressure.  The
stresses are proportional to the mean energy in the vertical field.

\begin{deluxetable}{cclrr@{.}ll}
\tablewidth{0pt}
\tablecaption{Simulations at location~A with weaker azimuthal
field.\label{tab:geometry}}
\tablehead{
\colhead{Label}    & \colhead{Name}    &
\colhead{$L/(2H)$} & \colhead{Zones}   &
\lbobz             & \colhead{\dratio} }
\startdata
R21  &Weaker-$B_y$  &$0.51$  &$128^2$&$-1$   &$0$ &$0.96$	\\
R23  &              &$0.051$ &$128^2$&$-2$   &$0$ &$0.18$	\\
R25  &              &$0.0051$&$128^2$&$-3$   &$0$ &$0.0027$	\\
\enddata
\tablecomments{The initial magnetic geometry is described in
\S~\ref{sec:geometry}.}
\end{deluxetable}

In the weaker azimuthal field calculations, as in the simulations~R1
through~R5 with the canonical magnetic geometry, the compression terms
$-(P+p)\divv$ are the major source of heating when magnetic pressure
in the turbulent stage is greater than gas pressure.  At a given ratio
of magnetic to gas pressure, the ratio of the net compression heating
to the vertical magnetic pressure increases with the mean strength of
the azimuthal field, indicating that the amount of heating depends on
the geometry of the field as well as its strength.

\subsection{Reduced Absorption Opacity
\label{sec:absorption}}

In radiation-dominated disks there may exist circumstances where the
time for exchange of energy between gas and radiation is similar to or
longer than the orbital period.  To examine how disequilibrium might
affect turbulence driven by the MRI, we carry out a version of the
canonical calculation with absorption opacity $10^6$ times lower, so
that $t_\mathrm{eqm}\approx 2\pi/\Omega_0$.  The initial conditions
and orbital frequency are identical to those used in the canonical
simulation.

The turbulent stage in this calculation lasts from 4.1~orbits to about
20~orbits.  During the most vigorous period of turbulence the
domain-averaged gas temperature climbs, reaching 5~times the initial
equilibrium temperature after 5.9~orbits.  In this period, heating of
the gas by $-p\divv$ and artificial viscosity is faster than the
cooling due to net emission of radiation.  After 8~orbits the gas
temperature declines, and from 15~orbits onwards it is close to the
radiation temperature.  Over the turbulent stage, the radiation
temperature increases gradually by a total of 6\%.  At its largest,
the domain-averaged gas pressure is 5.8\% of the initial radiation
pressure.  The smallest point-by-point ratio of radiation to gas
pressure is 1.5, and some parts of the flow are only marginally
radiation-dominated during the most vigorous turbulence.  The average
RMS density fluctuation from 2 to 6 orbits after the onset of
turbulence is 54\%, somewhat less than in the canonical calculation.
Gas temperature is almost uniform in the canonical run.  In the
version with reduced absorption opacity the RMS gas temperature
fluctuation is roughly equal to the density fluctuation from the onset
of turbulence until 12~orbits, and then decreases.  The RMS radiation
temperature fluctuation is near one-thirtieth of the density
fluctuation for the whole of the turbulent stage, as radiation
diffusion remains effective.

In similar low-absorption-opacity versions of the weaker-field
calculations~R3 and~R5, the domain-averaged gas temperature differs
little from the radiation temperature.  The time-averaged RMS gas
temperature fluctuations are intermediate between the gas temperature
fluctuations observed without radiation in runs N11-N20 and those in
the standard-opacity calculations R1-R5.

Overall, the lower absorption opacity leads to decoupling of gas and
radiation temperatures during the part of the turbulent stage when
heating is most rapid.  The density fluctuations, stress, and total
heating are little different from those in the versions of the
canonical calculation with the same initial magnetic fields.

\subsection{Ratio of Gas to Radiation Pressure
\label{sec:radius}}

The results of the earlier parts of this section indicate that
radiation effects may be important at locations where radiation
pressure greatly exceeds gas pressure.  In this final part we consider
whether radiation effects are significant when gas and radiation
pressures are comparable.  Under these conditions, magnetic pressure
is unlikely to be much larger than gas pressure.  A version of the
canonical calculation is carried out using the initial mass and energy
densities and orbital frequency found at location~B in the disk model
of \S~\ref{sec:ic}.  The domain size is unchanged at 51\% of the
thickness of the initial disk model.  Field strength is chosen so the
Alfv\'en speed is equal to the gas and radiation acoustic speeds, and
the pressures of the radiation, magnetic field, and gas are in the
ratio $1.25:0.833:1$.  The number of orbits required for photons to
diffuse vertically through the disk is independent of radius in the
radiation-dominated $\alpha$-prescription, so $L^2/D$ is 50~orbits
here as in the canonical calculation.

Time-averaged from 2 to 6 orbits after the start of turbulence, the
RMS density fluctuation is 12\%, an order of magnitude smaller than in
the canonical calculation.  The largest density on the grid during
this period is on average 2.3 times the smallest density.  The
time-averaged RMS temperature fluctuation is 0.7\%, about one-third
that in the canonical calculation.  The time-averaged RMS gas pressure
fluctuation is 85\% of the magnetic pressure fluctuation, suggesting
radiation pressure support of fluctuations is less important than in
the canonical simulation.

Gas plus radiation energy density increases between 2 and 6 orbits
after the start of turbulence by 5.2\% due to compression, and by
1.2\% due to artificial viscosity.  The compression heating is
comparable to the viscous dissipation assumed in the initial disk
model at location~B.  The dissipation in that model corresponds to an
increase by 8.1\% over the same time.  As in the canonical
calculation, the compression heating rate is largest in layers near
the boundaries between dense and underdense regions.  Overall, the
density and temperature fluctuations are weaker under these conditions
than at location~A, and the compression heating rate in units of
$(E+e) \Omega_0$ is about half as great.  The time-averaged total
accretion stress is $0.61$~times the viscous stress $\alpha(P+p)$
assumed in the initial disk model at location~B.  This ratio is
several times lower than observed in the canonical run at location~A.

In additional calculations with the field strength and grid spacing
both reduced, by factors of 10 in one case and 100 in the other, the
turbulence is almost exactly isothermal and incompressible, and no
compression heating is detected.  The three location~B calculations
are listed in table~\ref{tab:radius}.

\begin{deluxetable}{cclrr@{.}ll}
\tablewidth{0pt}
\tablecaption{Simulations at location~B with strong azimuthal
field\label{tab:radius}}
\tablehead{
\colhead{Label}    & \colhead{Name}    &
\colhead{$L/(2H)$} & \colhead{Zones}   &
\lbobz             & \colhead{\dratio} }
\startdata
R31  &Location~B  &$0.51$  &$128^2$& $0$   &$0$ &$0.37$         \\
R33  &            &$0.051$ &$128^2$&$-1$   &$0$ &$0.0077$       \\
R35  &            &$0.0051$&$128^2$&$-2$   &$0$ &$6.9\times 10^{-5}$\\
\enddata
\tablecomments{The initial magnetic geometry is described in
\S~\ref{sec:canonical}.}
\end{deluxetable}

\section{DISCUSSION\label{sec:discussion}}

To understand the luminous output of accretion disks around compact
objects we would like to know, given a certain rate of delivery of
material to the outer edge of the radiation-dominated region, how the
gas, radiation, and magnetic pressures are distributed in the
interior.  Some processes which are likely to determine these
quantities are angular momentum transfer, magnetic dynamo action,
dissipation and buoyant losses of field, outflows, heating of the disk
material, and losses of radiation through the disk surfaces.

\subsection{Magnetic Gains and Losses}

Magnetic energy increases in the present calculations through
differential orbital motion, and decreases through work done on the
gas and through numerical dissipation involving grid-scale averaging
of opposing fields.  The results obtained in \S~\ref{sec:zeronet}
suggest that magnetic fields may influence the gas and radiation by
transporting angular momentum outwards and driving turbulence leading
to heating.  The gas and radiation are likely to alter the fields in
turn through effects of compressibility and buoyancy.  The
calculations performed here cannot be used to learn about the
evolution of this coupled system because the stratification is
neglected, and the axisymmetry restricts field amplification to a
period of a few orbits, which is less than the thermal timescale for
the disk.  However, the association described in
\S~\ref{sec:fluctuations} between overdense regions and radial
magnetic field of one sign raises the possibility of a negative
feedback.  Regions with net radial field tend to produce strong
azimuthal field through shear.  Fluctuations with the opposite radial
field are compressed in this background, and are more likely to sink
towards the midplane because of their greater density.  Over time,
such buoyancy effects might favor removal of the original net radial
field to the disk surface layers.

In radiation-dominated disks at locations where vertical gravity is
important, magnetic losses due to buoyancy might be enhanced by
diffusion of radiation from the surroundings into rising, expanding
magnetized regions \citep{parker75,sr84}.  The hard X-ray spectra of a
range of objects powered by accretion onto black holes indicate
Comptonization in hot coronae of low optical depth
\citep{zdziarski99}.  The coronae may be heated by dissipation of
magnetic fields \citep{mf00}.  Magnetized coronae have been observed
to form in simulations of a local patch of accretion disk extending
several scale heights above and below the midplane \citep{ms00}.

\subsection{Radiation Gains and Losses}

In the calculations of \S~\ref{sec:zeronet}, turbulence is driven on
scales near the characteristic MRI wavelength.  When magnetic pressure
is larger than gas pressure, the turbulence is highly compressible,
and some of its kinetic energy is converted to photon energy through a
radiative damping similar to the process described for linear MHD
waves by \citet{ak98}.  As in the linear analysis, larger or denser
regions are less compressible because radiation diffuses from them
more slowly.  In smaller or less-dense regions, diffusion makes
radiation pressure gradient forces ineffective.  Damping is strongest
in the simulations in layers near the boundaries between regions of
high and low density.  The layers have transverse lengths comparable
to the MRI wavelength, and thicknesses typically a few percent of the
scale height of the $\alpha$-disk model from which the initial
conditions were selected.  The combination of density and thickness is
such that radiation diffuses through the damping layers in about an
orbital period.

Dissipation of kinetic energy in the simulations follows three paths.
Work is done in compressing the radiation directly, and in compressing
the gas.  Kinetic energy is also converted to gas internal energy
through the artificial viscous heating used to numerically capture
shocks.  In the canonical calculation the rate at which $-P\divv$ work
is done on the radiation is much greater than the rate at which
$-p\divv$ work is done on the gas, since radiation pressure is
initially 125 times gas pressure.  The time- and domain-averaged rate
of artificial viscous heating is about 10\% of the sum of the two
compression heating rates.  The majority of the energy deposited in
the gas is converted to photons within a few timesteps.  For the
location~A initial state, thermal equilibrium requires that linear
increases in energy density are partitioned $1000:1$ between radiation
and gas.

In the interiors of real accretion disks, energy lost by dissipation
of magnetic fields is likely released as heat.  This effect is not
included in the present ideal-MHD calculations, where energy lost
through numerical dissipation of the field vanishes with no
corresponding heating.  In the canonical calculation between 2 and 6
orbits after turbulence begins, the decrease in magnetic energy is
less than one tenth the increase in radiation energy density due to
compression heating.  We conclude that radiative damping of
compressive motions could be an important heating mechanism in
radiation-dominated disks.

In the picture developed by \citet{ss73}, density varies hardly at all
between the midplane and the disk surfaces, and radiation is lost by
diffusion.  The results outlined in \S~\ref{sec:fluctuations} indicate
that if the magnetic field exceeds equipartition with the gas, and
radiation diffuses an MRI wavelength in about an orbit, much of the
disk material may be compressed into small dense regions by magnetic
activity.  Under these conditions radiation would likely be lost from
the disk at an increased rate.  The time- and domain-averaged
radiation diffusion rate $\left<\left<D\right>\right>/L^2$ during the
turbulent stage in the canonical calculation is about twice the rate
in the uniform initial condition.  The radiation flux in an atmosphere
with density inhomogeneities may exceed the Eddington flux, with
low-density regions pushed up by radiation while high-density regions
are pulled down by gravity \citep{shaviv98,begelman01}.  The clumpy
state of the canonical calculation during the turbulent stage differs
from the two-phase equilibrium proposed by \citet{krolik98} in having
densities continuously distributed around a single most-common value
(figure~\ref{fig:histogram}).  The clumps here are formed, confined,
and destroyed by the magnetic field, and have lifetimes around one
orbital period.  We have included neither the magnetic dissipation
heating and Compton cooling which might establish a long-lived hot
phase, nor the thermal conduction which might set the size of
cold-phase clumps.

Large-scale buoyancy effects, which may enhance the overall cooling
rate, do not appear in our calculations since the vertical component
of gravity is neglected.  Ordinary convection might prove as effective
as diffusion in cooling a radiation-dominated $\alpha$-disk
\citep{akts01}.  Although stratification is not included here, the
photon bubble instability \citep{gammie98} could be present locally
where turbulent accelerations have components parallel to the
radiation flux.  For sufficiently large fluxes, linear photon bubble
modes with wavelengths somewhat shorter than the MRI characteristic
wavelength can grow faster than the MRI \citep{bs01}.  However the
similarities in gas and magnetic fluctuations between the canonical
simulation R1 and the simulation without radiation having the same
initial gas pressure, N11, suggest the resolved wavelengths of the
photon bubble instability are not important in this instance.  It is
unclear how well convective and photon bubble modes grow under
sustained vertical gravity in a background of turbulence driven by the
MRI.

\section{CONCLUSIONS
\label{sec:conclusions}}

We performed axisymmetric radiation MHD simulations of the MRI in
local patches of a radiation-dominated accretion disk, neglecting the
vertical component of gravity.  On initially-uniform magnetic fields
with a strong azimuthal component, radiation diffusion reduced the
growth rates of the linear modes in detailed agreement with the
analysis by \citet{bs01}.  On initially vertical fields, the
non-linear development of the instability led to channel solutions in
which compression of the dense layers was hastened by diffusive loss
of radiation pressure support.  On magnetic fields with zero net
vertical flux, the non-linear development resulted in about ten orbits
of decaying turbulence.  Because these axisymmetric calculations
showed no sustained dynamo action, the results cannot be used to learn
the steady-state strength of the magnetic field.  When magnetic
pressure was bigger than gas pressure and the distance that radiation
diffused in an orbit was at least comparable to the characteristic MRI
wavelength, density contrasts as large as the ratio of magnetic to gas
pressure were driven by magnetic forces.  Overdense regions were
destroyed and reformed in the turbulence about once per orbit.
Diffusion of radiation out of the compressed regions led to partial
separation of gas and radiation.  In cases with the standard free-free
and electron scattering opacities, the gas remained in good thermal
contact with the radiation, and both were approximately isothermal.
The flow was heated by radiation damping of compressive turbulent
motions.  Heating was fastest in regions where the compression rate
and radiation diffusion rate were similar.  Angular momentum was
transported outwards by a total accretion stress approximately four
times the pressure in the vertical component of the magnetic field.
At fixed vertical field strength, the stress showed little dependence
on the gas and radiation pressures.  The accretion stresses and
heating rates in calculations with initial magnetic pressure
comparable to radiation pressure were similar to the viscous stresses
and dissipation rates assumed in the $\alpha$-disk model used to
select the initial conditions.  Strong clumping of the gas occurring
at such magnetic pressures might allow more rapid cooling of
radiation-dominated disks than in the commonly assumed case of uniform
density.

\begin{acknowledgments}
This work was supported by the United States Department of Energy
under grant DFG-0398-DP-00215, and benefited from our discussions with
E.\ Agol, S.\ Balbus, O.\ Blaes, R.\ Blandford, R.\ Bowers, J.\
Cannizzo, C.\ Gammie, J.\ Hawley, W.-T.\ Kim, J.\ Krolik, M.\ C.\
Miller, E.\ Ostriker, and A.\ Socrates.  A.\ Young kindly made
computer time available.
\end{acknowledgments}


\end{document}